\documentclass[pre,showpacs,twocolumn,floatfix,amsmath,amsfonts]{revtex4}

\usepackage{epsfig}
\usepackage{graphicx}
\usepackage{amsmath}
\usepackage{amssymb}
\usepackage{dcolumn, color}

\begin{document}

\title{High energy density in the collision of $N$ kinks in the $\phi^4$ model}

\author{Aliakbar Moradi Marjaneh$^{1}$}

\author{Danial Saadatmand$^{2}$}
\email{saadatmand.d@gmail.com}

\author{Kun Zhou$^{3}$}

\author{Sergey V. Dmitriev$^{4,5}$}

\author{Mohammad Ebrahim Zomorrodian$^{1}$}

\affiliation{ $^1$Department of Physics, Ferdowsi University of Mashhad,
91775-1436 Mashhad, Iran \\
$^2$ Department of Physics, University of Sistan and Baluchestan, Zahedan, Iran
\\
$^3$ School of Mechanical and Aerospace Engineering, Nanyang Technological
 University, 50 Nanyang Avenue, Singapore 639798, Singapore
\\
$^4$ Institute for Metals Superplasticity Problems RAS, Khalturin Street 39, 450001 Ufa , Russia
\\
$^5$ Research Laboratory for Mechanics of New Nanomaterials, Peter the Great St. Petersburg Polytechnical University, St. Petersburg 195251,
Russia
 }

\begin{abstract}
Recently for the sine-Gordon equation it has been established that during collisions of $N$ slow kinks maximal energy density increases as $N^2$. In this numerical study, the same scaling rule is established for the non-integrable $\phi^4$ model for $N\le 5$. For odd (even) $N$ the maximal energy density is in the form of potential (kinetic) energy density. Maximal elastic strain is also calculated. In addition, the effect of the kink's internal modes on the maximal energy density is analysed for $N=1$, 2, and 3. Our results suggest that in multi-soliton collisions very high energy density can be achieved in a controllable manner.
\end{abstract}
 \maketitle

\section{Introduction}
Solitons play an important role in a wide range of areas such as optics \cite{Optics1,Optics2}, superconducting Josephson junction arrays \cite{JJ1,JJ2,JJ3}, particle and nuclear physics \cite{Manton,Weigel}, condensed matter physics \cite{Bishop,Ferro1,Ferro2}, and many others \cite{Belova,BraunKivshar,BookSGE,nSGE}. The $\phi^4$ model, which is a particular case of the Klein-Gordon equation, was extensively studied in relation to the resonant kink-antikink scattering \cite{f1,f2,f3,f4}, kinks interaction with defects \cite{Belova,FeiKivshar,Danialphi4a,Danialphi4b,DanialSGE}, with the periodically modulated on-site potential \cite{FeiKonotop}, and with ac external force \cite{Quintero}. The effect of discretization scheme on the properties of the $\phi^4$ kinks has been analysed in \cite{Barashenkov,Avadh,Ishani}. Solitons in nonintegrable models can support internal vibrational modes \cite{KPCP} which make soliton dynamics much richer than in the integrable sine-Gordon equation \cite{BraunKivshar,BookSGE,Quintero,phi6a,phi6b,phi8,Khare}, since energy exchange between translational and vibrational modes is possible.

Solitary waves are very robust with respect to small perturbations, they can travel long distances and survive collisions with each other and thus, they are very efficient in energy transfer. Colliding solitons can produce high energy density spots and for many applications it is important to know how large the energy density can be in multi-soliton collisions. Recently we have addressed this issue in the realm of the integrable sine-Gordon equation \cite{DanialPRD}. It was found that maximal energy density that can be observed in collision of $N$ slowly moving kinks/antikinks is proportional to $N^2$, while total energy of the system is proportional to $N$. Such a high energy density can be achieved only if the kinks and antikinks approach the collision point in an alternating array, where each soliton has nearest neighbors of the opposite topological charge and thus, all solitons attract their nearest neighbors. Interestingly, when $N$ is odd (even) the maxiaml energy density is in the form of potential (kinetic) energy with kinetic (potential) energy being zero. In the present study the same problem is addressed for the $\phi^4$ model to see the effect of non-integrability of the model and the effect of the kink's internal modes (IM) on the maximal energy density and on the scenarios of multi-kink collisions.

The paper is organized as follows. In Sec. \ref{Sec:II} the problem to solve is described. In Sec. \ref{Sec:III}, by integrating numerically the $\phi^4$ equation of motion, we estimate the maximal energy density observed in the collision of $N$ slowly moving kinks and antikinks with no initially excited kink's IM. The effect of the initially excited kink's IM on the maximal energy density is discussed in Sec. \ref{Sec:IV} for $N=1$, 2, and 3. The key results of the present study are summarized in Sec.~\ref{Sec:V}.

\section {General statements} \label{Sec:II}
We consider the following $\phi^4$ equation in (1+1) dimension
\begin{equation}\label{Phi4}
\phi_{tt} - \phi_{xx} - 2\phi(1-\phi^2) = 0,
\end{equation}
where $\phi(x,t)$ is the scalar field and lower indices indicate partial differentiation. The total energy of the field is given by
\begin{equation}\label{Energy}
U=K+E+P=\frac{1}{2}\int\limits_{-\infty}^{\infty} \Big[ \phi_t^2 +\phi_x^2 +(1-\phi^2)^2 \Big]dx,\,
\end{equation}
where the first to the third terms in the right-hand side give the kinetic energy, $K$, the elastic strain energy, $E$, and the on-site potential energy, $P$, respectively. The corresponding integrands describe the three contributions to the total energy density of the $\phi^4$ field,
\begin{eqnarray}\label{Edensity}
u(x,t)&=& k(x,t)+e(x,t)+p(x,t) \nonumber \\ &=&\frac{1}{2}\phi_t^2+ \frac{1}{2}\phi_x^2+\frac{1}{2}(1-\phi^2)^2.
\end{eqnarray}
Elastic strain is defined as follows
\begin{eqnarray}\label{Strain}
\varepsilon(x,t)=\phi_x,
\end{eqnarray}
which is positive (negative) for tension (compression).

The exact solitary wave solution to the $\phi^4$ field Eq.~(\ref{Phi4}) is
the kink (antikink)
\begin{equation}\label{Kink}
   \phi(x,t)=\pm \tanh [\delta (x-V t)],
\end{equation}
where $V$ is the kink velocity and $\delta =1/\sqrt{1 - V^2}$ is the kink inverse width. The upper (lower) sign in Eq.~(\ref{Kink}) corresponds to the kink
(antikink). Substituting Eq.~(\ref{Kink}) into Eq.~(\ref{Energy})
one finds the total energy of the kink
\begin{equation}\label{KBEnergy}
   U=\frac{4\delta}{3}.
\end{equation}
For the non-integrable $\phi^4$ equation exact $N$-soliton solutions are not known. However, it is obvious that the total energy of $N$ non-overlapping kinks/antikinks is equal to $NU$.

In this study, we will consider only slow solitons ($|V| \ll 1$) so that $\delta \approx 1$. Then, we can write approximately that for one slow kink $U \approx 4/3$, and for $N$ slow kinks/antikinks total energy is about $4N/3$.

To perform numerical simulations, the discrete version of
Eq.~(\ref{Phi4}) is proposed as follows
\begin{eqnarray}\label{Phi4discrete}
&&\frac{d^2\phi_n}{dt^2} - \frac{1}{h^2}(\phi_{n-1} -2\phi_{n}+\phi_{n+1}) \nonumber \\
&&+\frac{1}{12h^2}(\phi_{n-2}-4\phi_{n-1} +6\phi_{n}-4\phi_{n+1}+\phi_{n+2})\nonumber \\
&&-2\phi_n(1-\phi_n^2)  = 0,
\end{eqnarray}
where $h$ is the lattice spacing, $n=0,\pm1,\pm2,...$, and $\phi_n(t)=\phi(nh,t)$. In order to minimize the effect of discreteness, the term $\phi_{xx}$ in Eq.~(\ref{Phi4discrete}) is discretized with the accuracy $O(h^4)$ \cite{BraunKivshar,DanialPRD}. The equations of motion in the form of Eq.~(\ref{Phi4discrete}) were integrated with respect to the time using an explicit scheme with the time step $\tau$ and the accuracy of $O(\tau^4)$. The simulations reported here were carried out for $h = 0.1$, $h = 0.05$ and $\tau = 0.005$.

For the sine-Gordon equation it has been shown that $N$ kinks/antikinks can collide at one point if each of them has nearest neighbors with the opposite topological charge, and thus all quasiparticles are mutually attractive \cite{DanialPRD,PREcollisions}. Here we also set initial conditions with the help of the exact single-soliton solution Eq. (\ref{Kink}) in a way that the solitons initially do not overlap and have alternating topological charges. Solitons in the middle of the array have smaller velocities, while those further from the middle have larger absolute values of the velocities chosen such that all of them collide at one point.

\section {Maximal energy density and strain in the collision of $N$ kinks and antikinks} \label{Sec:III}

In this Section, we first give the exacts values of the maxiamal energy densities and maximal elastic strain for the single standing kink. Then the same quantities are calculated numerically for $N$ slowly moving kinks and antikinks colliding at one point ($N=2$, 3, 4, and 5).

\subsection{Case $N=1$} \label{Sec:N1}

Substituting Eq.~(\ref{Kink}) with $V=0$ into Eq.~(\ref{Edensity}) the maximal values of the energy densities of standing kink or antikink can be found as
\begin{equation}\label{umax1}
   u^{(1)}_{\max}=1,
\end{equation}
\begin{equation}\label{kmax1}
   k^{(1)}_{\max}=0,
\end{equation}
\begin{equation}\label{emax1}
   e^{(1)}_{\max}=1/2,
\end{equation}
\begin{equation}\label{pmax1}
   p^{(1)}_{\max}=1/2.
\end{equation}
For $N=1$ total maximal energy density, $u^{(1)}_{\max}$, is equal to the sum of $k^{(1)}_{\max}$, $e^{(1)}_{\max}$, and $p^{(1)}_{\max}$. It will be seen later that this is not so for $N>1$.

By substituting Eq.~(\ref{Kink}) with $V=0$ into Eq.~(\ref{Strain}) one finds the extreme values of strain as
\begin{equation}\label{epsmax1}
   \varepsilon^{(1)}_{\max}= 1, \quad \varepsilon^{(1)}_{\min}=-1,
\end{equation}
where the first (second) result is for the kink (antikink).

\subsection{Case $N=2$} \label{Sec:N2}

\begin{figure}\center
\includegraphics[width=7.5cm]{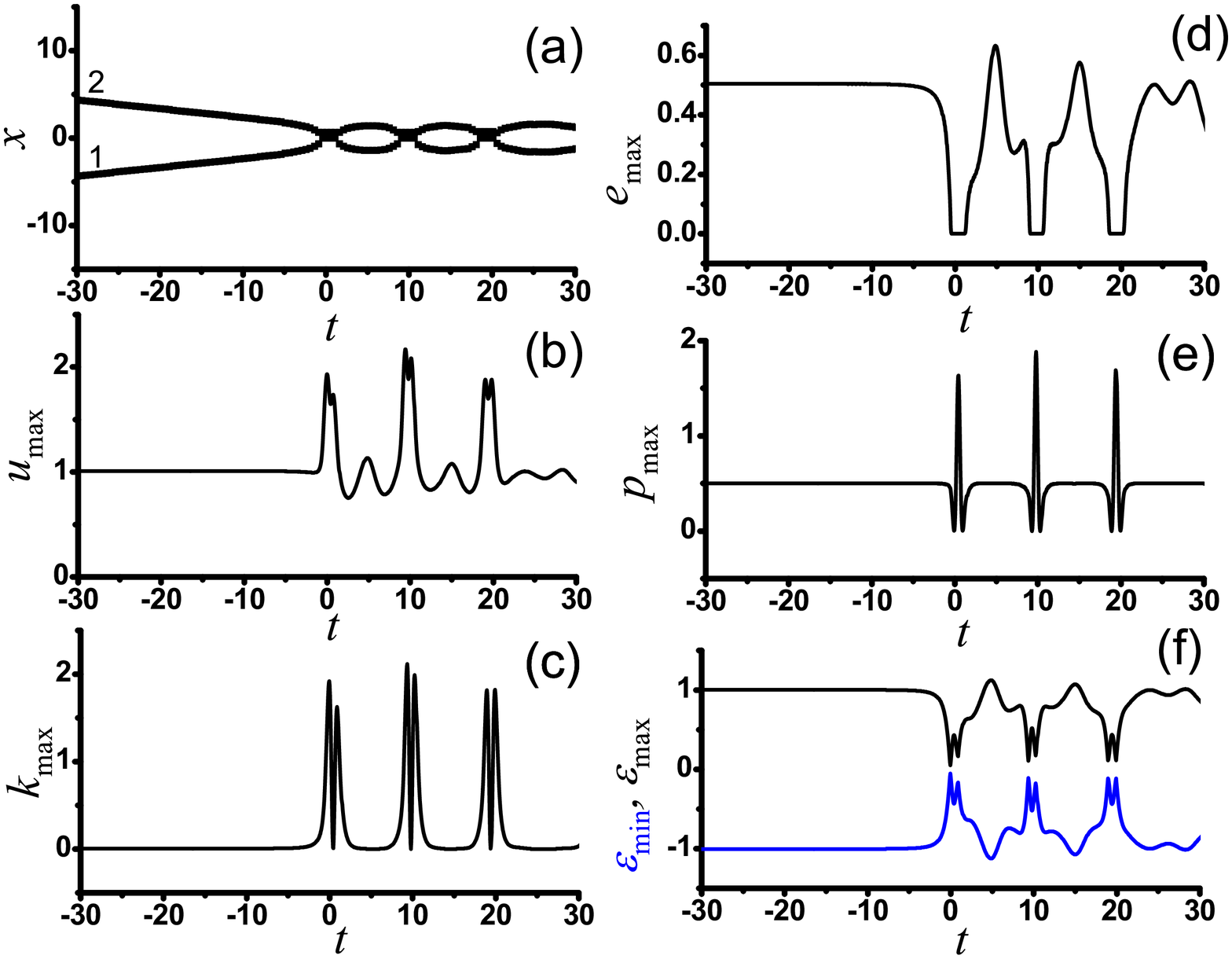}
\caption{Results for the collision of kink and antikink with the initial positions $x_1=-10$, $x_2=10$, and the initial velocities $V_1=0.1$, $V_2=-0.1$. (a)~Trajectories of the soliton cores are shown by plotting the regions of the $(t,x)$-plane with the total energy density $u>1/2$, for the time domain close to the collision point. (b-f)~Maximal over spatial coordinate $u$, $k$, $e$, $p$, and $\varepsilon$ as the functions of time, respectively.}\label{fig1}
\end{figure}

Kink-antikink collision are simulated by setting initial conditions with the help of Eq.~(\ref{Kink}). Initially the kink and antikink do not overlap having positions $x_1=-10$ and $x_2=10$. Their initial velocities are $V_1=0.1$ and $V_2=-0.1$. The results are presented in Fig.~\ref{fig1} for the time domain close to the collision point. In (a) the regions of the ($t,x$)-plane with the total energy density $u>1/2$ are shown to reveal the cores of the solitons. In (b) to (f), as the functions of time, shown are the maximal over $x$ values of $u$, $k$, $e$, $p$, and $\varepsilon$, respectively. From (a) it can be seen that a bound state of the kink and antikink is formed and they collide many times. From (b) to (f) it follows that in the first collision the maximal values of the energy densities and extreme values of the elastic strain are
\begin{equation}\label{umax2}
   u^{(2)}_{\max}\approx 2,
\end{equation}
\begin{equation}\label{kmax2}
   k^{(2)}_{\max}\approx 2,
\end{equation}
\begin{equation}\label{emax2}
   e^{(2)}_{\max}\approx 0.5,
\end{equation}
\begin{equation}\label{pmax2}
   p^{(2)}_{\max}\approx 1.75,
\end{equation}
\begin{equation}\label{epsmax2}
   \varepsilon^{(2)}_{\max}\approx 1, \quad \varepsilon^{(2)}_{\min}\approx -1.
\end{equation}
It is clear that $k^{(2)}_{\max}+e^{(2)}_{\max}+p^{(2)}_{\max}>u^{(2)}_{\max}$. This can be explained by careful observation of Fig.~\ref{fig1}. From (c) it can be seen that during a kink-antikink collision maximal over $x$ kinetic energy density has two sharp peaks, while from (e) it is seen that in between these peaks $p_{\max}(t)$ has a sharp peak. From (d) it follows that $e_{\max}(t)$ is nearly zero during the kink-antikink collision and its maximal value is observed for the well separated solitons. Thus, maximal values of $k^{(2)}_{\max}$, $e^{(2)}_{\max}$, and $p^{(2)}_{\max}$ are observed at different times and that is why their sum is greater than $u^{(2)}_{\max}$.

We note that the above results depend on the lattice spacing used in the simulations. Particularly, the maximal total energy density in the first collision is found to be $u^{(2)}_{\max}\approx 2.22$ for $h=0.1$ and $u^{(2)}_{\max}\approx 2.16$ for $h=0.05$. The result for smaller $h$ is more accurate because the discreteness effect is smaller in this case.

Another important comment is that the maximal values of the analysed quantities observed in the subsequent collisions can be greater than in the first collision, in spite of the fact that in each collision the kink-antikink pair radiates a portion of energy in the form of small-amplitude running waves. This can be explained by the kink's internal modes that are excited during the first collision. The effect of the kink's internal modes will be addressed in more detail in Sec.~\ref{Sec:IV}.

\subsection{Case $N=3$} \label{Sec:N3}

\begin{figure}\center
\includegraphics[width=7.5cm]{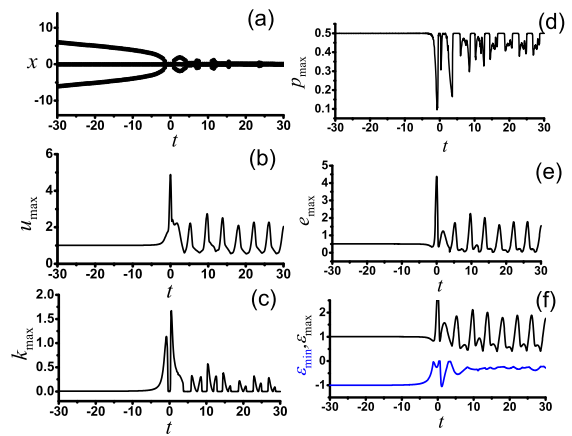}
\caption{Same as in Fig.~\ref{fig1} but for the kink-antikink-kink collision. Initial soliton positions and velocities are $x_1=-20$, $x_2=0$, $x_3=20$, $V_1=0.1$, $V_2=0$, and $V_3=-0.1$.}\label{fig2}
\end{figure}

Kink-antikink-kink collision is simulated for the initial soliton positions $x_1=-20$, $x_2=0$, and $x_3=20$, and velocities $V_1=0.1$, $V_2=0$, and $V_3=-0.1$. The results are shown in Fig.~\ref{fig2} similarly with the case of $N=2$. The maximal values of the energy densities and tensile and compressive strain are
\begin{equation}\label{umax3}
   u^{(3)}_{\max}\approx 5,
\end{equation}
\begin{equation}\label{kmax3}
   k^{(3)}_{\max}\approx 1.7,
\end{equation}
\begin{equation}\label{emax3}
   e^{(3)}_{\max}\approx 4.5,
\end{equation}
\begin{equation}\label{pmax3}
   p^{(3)}_{\max}\approx 0.5,
\end{equation}
\begin{equation}\label{epsmax3}
   \varepsilon^{(3)}_{\max}\approx 2.5, \quad \varepsilon^{(3)}_{\min}\approx -1.
\end{equation}
More precisely, $u^{(3)}_{\max}=4.82$ for $h=0.1$ and $u^{(3)}_{\max}=4.91$ for $h=0.05$.

\subsection{Case $N=4$} \label{Sec:N4}

\begin{figure}\center
\includegraphics[width=7.5cm]{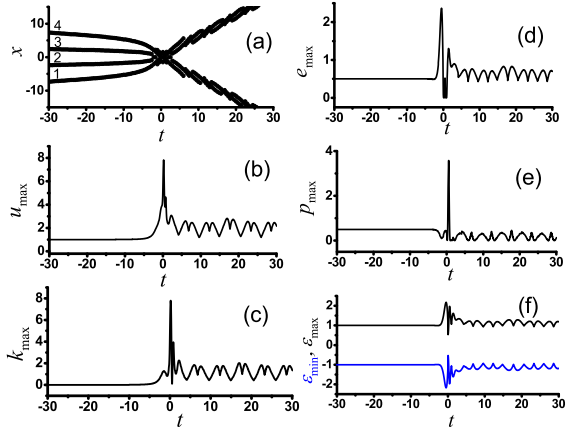}
\caption{Same as in Fig.~\ref{fig1} but for the collision of four kinks/antikinks. Initial conditions are set with $x_1=-x_4=-22.276$, $x_2=-x_3=-10$ and $V_1=-V_4=0.05$, $V_2=-V_3=0.025$.} \label{fig3}
\end{figure}

In the initial configuration shown in Fig.~\ref{fig3}~(a), solitons 1 and 3 are
the kinks, while 2 and 4 are the antikinks. They collide at one
point provided that their initial coordinates and velocities
are chosen as follows: $x_1=-x_4=-22.276$, $V_1=-V_4=0.05$, $x_2=-x_3=-10$ and $V_2=-V_3=0.025$.

It can be seen in Fig.~\ref{fig3}~(b-f) that at the point of collision of the four
kinks
\begin{equation}\label{umax4}
   u^{(4)}_{\max}\approx 8,
\end{equation}
\begin{equation}\label{kmax4}
   k^{(4)}_{\max}\approx 8,
\end{equation}
\begin{equation}\label{emax4}
   e^{(4)}_{\max}\approx 2.4,
\end{equation}
\begin{equation}\label{pmax4}
   p^{(4)}_{\max}\approx 3.5,
\end{equation}
\begin{equation}\label{epsmax4}
   \varepsilon^{(4)}_{\max}\approx 2.2, \quad \varepsilon^{(4)}_{\min}\approx -2.2.
\end{equation}
More precisely, for $h=0.1$ the largest energy density we could obtain by varying the parameter $x_{1}=-x_{4}$ was $u^{(4)}_{\max}=8.44$, while for $h=0.05$ it was $u^{(4)}_{\max}=8.26$.

\subsection{Case $N=5$} \label{Sec:N5}

\begin{figure}\center
\includegraphics[width=7.5cm]{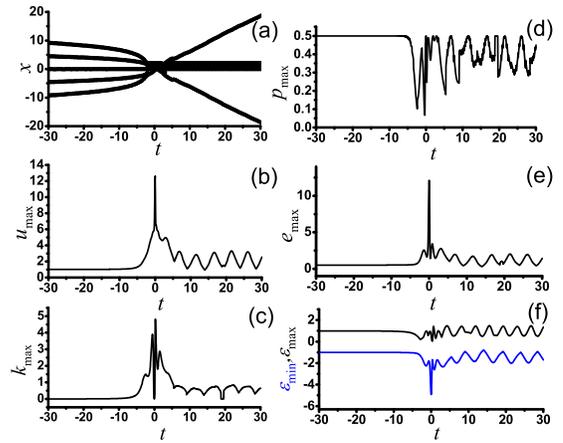}
\caption{Same as in Fig.~\ref{fig1} but for the collision of five kinks/antikinks. Initial soliton positions are $x_1=-x_5=-28.192867$, $x_2=-x_4=-14.0$, and $x_3=0$, while their initial velocities are $V_1=-V_5=0.05$, $V_2=-V_4=0.025$, and $V_3=0$.
}\label{fig4}
\end{figure}

In the initial configuration shown in Fig.~\ref{fig4}~(a), solitons 1, 3,
and 5 are kinks and 2 and 4 are antikinks. The kink 3 is located at the
origin and initially it is at rest, $x_3=0$ and $V_3=0$. The antikinks 2 and 4 have
velocities $V_2=-V_4=0.025$ and initial positions $x_2=-x_4=-14.0$. By symmetry
the solitons 2, 3, and 4 collide at one point. For the kinks 1 and 5 we take
two times larger velocities $V_1=-V_5=0.05$ and choose their initial coordinates
to achieve the collision of five solitons at one point. This happens for
$x_1=-x_5=-28.192867$. As it can be seen from Fig.~\ref{fig4}(b-f), when five solitons collide,
\begin{equation}\label{umax5}
   u^{(5)}_{\max}\approx 13,
\end{equation}
\begin{equation}\label{kmax5}
   k^{(5)}_{\max}\approx 5,
\end{equation}
\begin{equation}\label{emax5}
   e^{(5)}_{\max}\approx 12.5,
\end{equation}
\begin{equation}\label{pmax5}
   p^{(5)}_{\max}\approx 0.5,
\end{equation}
\begin{equation}\label{epsmax5}
   \varepsilon^{(5)}_{\max}\approx 1.5, \quad \varepsilon^{(5)}_{\min}\approx -5.
\end{equation}
More precisely, the maximal energy density is $u^{(5)}_{\max}=10.72$ for $h=0.1$ and $u^{(5)}_{\max}=12.85$ for $h=0.05$.

The results of this Section are collected in Table \ref{cnls_numer_summary}.

\begin{table}[pht] \center \begin{small}
\caption{Summary on maximal energy densities and tensile and compressive elastic strains in collision of $N$ solitons.} \centering
\begin{tabular}{ | c | c | c |c | c | c | }
\hline
$N$                  & $1$   & $2$   & $3$   & $4$   & $5$      \\
$u_{\max}$           & $1$   & $2$   & $5$   & $8$   & $13$ \\
$k_{\max}$           & $0$   & $2$   & $1.7$ & $8$   & $5$  \\
$e_{\max}$           & $1/2$ & $0.5$ & $4.5$ & $2.4$ & $12.5$  \\
$p_{\max}$           & $1/2$ & $1.75$& $0.5$ & $3.5$ & $0.5$  \\
$\varepsilon_{\max}$ & $1$   & $1$   & $2.5$ & $2.2$ & $1.5$  \\
$\varepsilon_{\min}$ & $-1$  & $-1$  & $-1$  & $-2.2$& $-5$  \\
 \hline
\end{tabular}
\label{cnls_numer_summary}
\end{small}
\end{table}

\section {Effect of kink's internal mode} \label{Sec:IV}

In this Section, we consider collisions of the kinks bearing IM taking for the lattice spacing $h=0.005$. An approximate solution to Eq.~(\ref{Phi4}), which leads to the kink with the IM having amplitude $A$ and frequency $\omega=\sqrt{3}$, is considered as follows \cite{Belova}
\begin{equation}\label{IM}
   \Phi(x,t)=\phi(x,t)+A\eta(x,t),
\end{equation}
in which $\eta(x,t)$ describes the kink's IM profile in the following form
\begin{equation}\label{Profile}
   \eta(x,t)=\sqrt{\frac{3}{2}}\tanh[\delta(x-Vt)]{\rm sech}[\delta(x-Vt)].
\end{equation}

\begin{figure}[t]\center
\includegraphics[width=7.3cm]{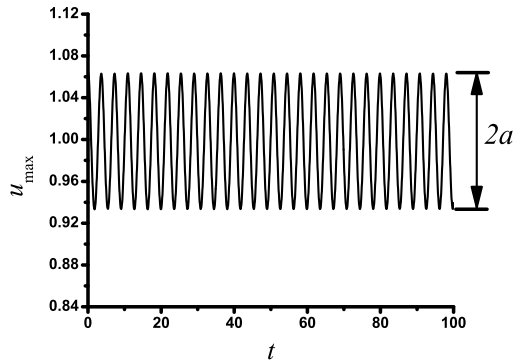}
\caption{Time dependence of the maximal energy density for single standing kink bearing IM of amplitude $A=0.05$. $u_{\max}(t)$ oscillates with the amplitude $a$ and frequency $\omega \approx \sqrt{3}$, which is the IM frequency.}
\label{fig5}
\end{figure}

Firstly we excite IM with the amplitude $A=0.05$ on a standing kink and calculate maximal over $x$ total energy as the function of time, $u_{\max}(t)$. The result is presented in Fig.~\ref{fig5}. It can be seen that $u_{\max}$ oscillates with the period $T=3.64$ and frequency $\omega=2\pi/T \approx \sqrt{3}$ about the value $u^{(1)}_{\max}=1$. Let us denote the amplitude of oscillation of $u_{\max}$ as $a$. In this example $a=0.062$. In Fig.~\ref{fig7}~(a) the amplitude $a$ is plotted as the function of the IM amplitude $A$ to reveal the linear relation between them.

\begin{figure}[t]\center
\includegraphics[width=7.3cm]{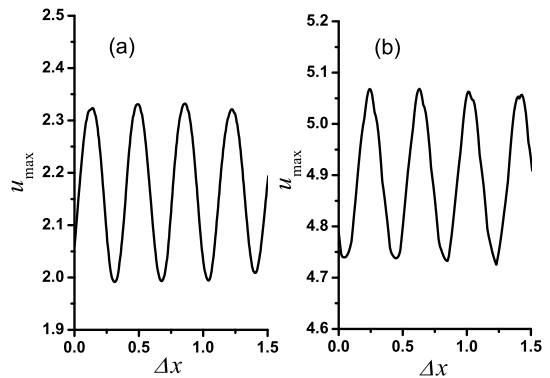}
\caption{Effect of kink's IM on the maximal energy density
in (a) two-kink and (b) three-kink collisions. Amplitude of the kink's IM is $A=0.05$ in both cases.}
\label{fig6}
\end{figure}

\begin{figure}[t]\center
\includegraphics[width=7.3cm]{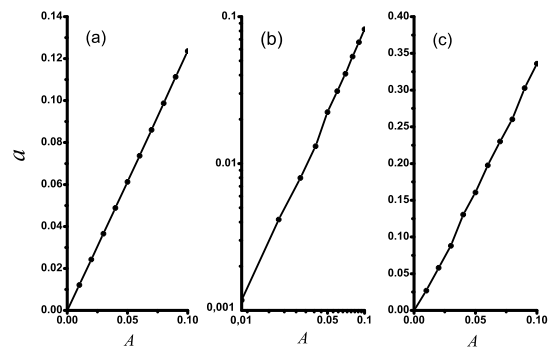}
\caption{Amplitude of maximal energy density $a$ as the function of kink's IM amplitude $A$ in (a) one-kink, (b) two-kink and (c) three-kinks collisions. The lattice spacing is $h=0.05$ in three cases. Note the use of the double logarithmic scale in (b) and linear scales in (a) and (c).}
\label{fig7}
\end{figure}

Secondly we consider the symmetric collisions of kink and antikink each bearing IM of amplitude $A=0.05$. Solitons' initial velocities are $V_{1,2}=\pm 0.1$ and initial positions are $x_{1,2}=\mp 10\pm\Delta x$, where parameter $\Delta x$ is introduced to study the effect of IM phase on the collision. In Fig.~\ref{fig6}~(a) the maximal energy density in collision of two kinks bearing IM is shown as the function of $\Delta x$. This function oscillates near the value $u^{(2)}_{\max}=2.16$ with the period $V_1T=0.364$, where $T=3.64$ is the IM oscillation period. The oscillation amplitude in this example is $a=0.17$. In Fig.~\ref{fig7}~(b) the amplitude $a$ is plotted as the function of the IM amplitude $A$ in the log-log scale to demonstrate the quadratic relation between them, $a\sim A^2$.

Finally, we study symmetric kink-antikink-kink collisions with two kinks bearing IM of amplitude $A=0.05$ and antikink free of IM. The solitons' initial positions are $x_{1}=-20+\Delta x$, $x_2=0$, and $x_{3}=20-\Delta x$, while velocities are $V_{1}=0.1$, $V_2=0$, and $V_{3}=-0.1$. Maximal total energy observed in the collisions, $u_{\max}$, is shown in Fig.~\ref{fig6}~(b) as the function of $\Delta x$. This function oscillates near the value $u^{(3)}_{\max}=4.91$ with the amplitude $a=0.165$ and with the period $V_1T=0.364$, where $T=3.64$ is the IM oscillation period. In Fig.~\ref{fig7}~(c) the amplitude $a$ is plotted as the function of the IM amplitude $A$ to show the linear relation between them.


\section {Conclusions and future challenges} \label{Sec:V}

In this paper, the maximal energy densities and extreme values of elastic strain in the collision of $N$ slow kinks/antikinks (with $N \leq 5$) in the non-integrable $\phi^4$ model were calculated numerically. Our results are collected in Table~\ref{cnls_numer_summary}.

We conclude that the maximal total energy density that can be achieved in collision of $N$ slow kinks/antikinks in the $\phi^4$ model is equal to
\begin{eqnarray}\label{fit}
   u_{\max}^{(N)}\approx \frac{N^2}{2} \quad &&{\rm for \,\, even }\,\, N, \nonumber \\
   u_{\max}^{(N)}\approx \frac{N^2+1}{2} \quad &&{\rm for \,\, odd }\,\, N.
\end{eqnarray}
The same rule was found earlier for the sine-Gordon equation \cite{DanialPRD}.

These values of the maximal total energy density can be achieved when all $N$ kinks/antikinks collide at one point. This happens when the kinks and antikinks approach the collision point alternatively (i.e., no two adjacent solitons are of the same type). When arranged in this way, the solitons attract each other and their cores can merge producing a controllably high energy density spots, as we have demonstrated herein.

We have separated the total energy density, $u$, into three components, the kinetic energy density, $k$, the elastic strain energy density, $e$, and the potential energy density due to the on-site potential, $p$. Their maximal values observed in the collisions of $N$ kinks/antikinks are also given in Table~\ref{cnls_numer_summary}. We note that $k^{(N)}_{\max}$ increases rapidly with $N$ for even $N$, while for odd $N$ a rapid growth with $N$ is observed for $e^{(N)}_{\max}$. These two energy densities have a dominant contribution to the maximal total energy density.

For many applications, e.g., in the solid state physics, it is important to know the maximal values and the sign of the maximal elastic strain observed in $N$-soliton collisions. The last two lines of Table~\ref{cnls_numer_summary} contain this information. For $N=3$ the maximal tensile strain of $2.5$ is registered, which is 2.5 times larger than in the core of a single kink. For $N=5$, maximal compressive strain of $-5$ is observed, which is 5 times larger than in the core of an antikink. 

In Sec. \ref{Sec:IV}, the effect of kink's IM on the maximal total energy density was studied for single standing kink, for the kink-antikink collisions, and for the kink-antikink-kink collisions. The results presented in Fig.~\ref{fig7} can be summarized as follows. The variation of total energy density, $u_{\max}$, due to the kink's IM increases linearly with IM amplitude, $A$, in the cases of single-kink and kink-antikink-kink collisions, while for the kink-antikink collisions it increases with $A$ quadratically. For the maximal studied IM amplitude of $A=0.1$ the maximal increase in $u^{(1)}_{\max}$ due to the excitation of IM is 12~\%, in $u^{(2)}_{\max}$ it is 5~\%, and in $u^{(3)}_{\max}$ it is 7~\%.

For the future works, it is important to calculate the maximal energy density that can be achieved in multi-soliton collisions in other integrable and non-integrable systems of different dimensionality. For example, one can examine similar issues and design such collisions in other Klein-Gordon field theoretical models (e.g. in the $\phi^6$, $\phi^8$, or $\phi^{12}$ models~\cite{phi6a,phi6b,phi8,Khare}), as well as in the nonlinear Schr{\"o}dinger equation. It would be particularly interesting to explore if the relevant phenomenology persists therein. Next, it would be extremely interesting to search for the physical phenomena that can be related to the high energy density spots or/and highly strained regions generated by multi-soliton collisions. Some of these ideas are under consideration and will be reported in the future works.

\section*{Acknowledgments}

D.S. and A.M.M. thanks the financial support of the Institute for Metals Superplasticity Problems, Ufa, Russia. S.V.D. thanks financial support provided by the Russian Science
Foundation grant 16-12-10175.


\begin{thebibliography}{99}

%\vspace{-15mm}

\bibitem{Optics1}
Yu.~S.~Kivshar, G.~P. Agrawal, Optical Solitons. From Fibers to Photonic Crystals, Academic Press, Burlington (2003).

\bibitem{Optics2}
 S.~V.~Suchkov,  A.~A.~Sukhorukov,  J.~Huang,  S.~V.~Dmitriev,  C.~Lee and Yu.~S.~Kivshar, Laser Photonics Rev. {\bf 10}, 177213 (2016).

\bibitem{JJ1}
P.~S.~Lomdahl, J. Stat. Phys. {\bf 39}, 551 (1985).

\bibitem{JJ2}
J.~Pfeiffer, M.~Schuster, A.~A.~Abdumalikov, and A.~V.~Ustinov, Phys. Rev. Lett. {\bf 96}, 034103 (2006).

\bibitem{JJ3} S. Watanabe, H.~S.~J. van der Zant, S.~H.~Strogatz, T.~P.~Orlando, Physica D {\bf 97}, 429 (1996).

\bibitem{Manton} N.~Manton and P.~Sutcliffe, Topological Solitons, Cambridge University Press, Cambridge, (UK), (2004).

\bibitem{Weigel} H.~Weigel, Chiral Soliton Models for Baryons, Lect. Notes Phys. 743 (2008) 1.

\bibitem{Bishop} A.~R. Bishop and T. Schneider (Eds.), Solitons in Condensed Matter Physics, Springer Verlag, Berlin (1978).

\bibitem{Ferro1} S.~V.~Dmitriev, K.~Abe, T.~Shigenari, Phys. Rev. B {\bf 58}, 2513 (1998).

\bibitem{Ferro2} S.~V.~Dmitriev, K.~Abe, T.~Shigenari, J. Phys. Soc. Jpn {\bf 65}, 3938 (1996).

\bibitem{Belova} T.~I. Belova, A.~E. Kudryavtsev, Phys. Usp. {\bf 40}, 359 (1997).

\bibitem{BraunKivshar} O.~M. Braun, Yu.S. Kivshar, \textit{The Frenkel-Kontorova Model:
Concepts, Methods, and Applications} (Springer, Berlin, 2004).

\bibitem{BookSGE} J.~Cuevas-Maraver, P.~G. Kevrekidis, and F.~Williams, eds.,
The sine-Gordon Model and its Applications. From Pendula and Josephson
Junctions to Gravity and High Energy Physics, Springer, Berlin, 2014.

\bibitem{nSGE} S.~Y. Lou, H.~C. Hu, and X.~Y. Tang, Phys. Rev. E {\bf 71}, 036604 (2005).

\bibitem{f1}
R.~H.~Goodman, R.~Haberman. Phys. Rev. Lett. {\bf 98}, 104103 (2007).

\bibitem{f2}
D.~K.~Campbell, J.~S.~Schonfeld, and C.~A.~Wingate, Physica D {\bf 9}, 1 (1983);

\bibitem{f3}
D.~K.~Campbell and M.~Peyrard, Physica D {\bf 18}, 47 (1986).

\bibitem{f4}
D.~K.~Campbell and M. Peyrard, Physica D {\bf 19}, 165 (1986).

\bibitem{FeiKivshar}
Z. Fei, Yu.~S.~Kivshar, and L.~Vazquez, Phys. Rev. A {\bf 46}, 5214 (1992).

\bibitem{Danialphi4a}
D.~Saadatmand, S.~Dmitriev, D.~Borisov, P.~Kevrekidis, M.~Fatykhov, K.~Javidan, JETP Lett. {\bf 101}, 497 (2015).

\bibitem{Danialphi4b}
D. Saadatmand, S.~V.~Dmitriev, D.~I.~Borisov, P.~G.~Kevrekidis, M.~A.~Fatykhov, and K.~Javidan, Commun. Nonlinear Sci. Numer. Simulat. {\bf 29}, 267 (2015).

\bibitem{DanialSGE} D.~Saadatmand, S.~V.~Dmitriev, D.~I.~Borisov, and P.~G.~Kevrekidis, Phys. Rev. E {\bf 90}, 052902 (2014).

\bibitem{JavidanPRE} E.~Hakimi, and K.~Javidan, Phys. Rev. E {\bf 80}, 016606 (2009).

\bibitem{FeiKonotop} Z.~Fei, V.~V.~Konotop, M.~Peyrard, and L.~Vazquez, Phys. Rev. E {\bf 48}, 548 (1993).

\bibitem{Quintero} N.~R.~Quintero, A.~Sanchez, and F.~G.~Mertens, Phys. Rev. E {\bf 62}, 5695 (2000).

\bibitem{Barashenkov}
I.~V.~Barashenkov, O.~F.~Oxtoby and D.~E.~Pelinovsky, Phys. Rev. E {\bf 72}, 35602R (2005).

\bibitem{Avadh}
S.~V.~Dmitriev, P.~G.~Kevrekidis, A.~Khare, and A.~Saxena, J. Phys. A: Math. Theor. {\bf 40}, 6267 (2007).

\bibitem{Ishani}
I.~Roy, S.~V.~Dmitriev, P.~G.~Kevrekidis, and A.~Saxena, Phys. Rev. E {\bf 76}, 026601 (2007).

\bibitem{KPCP} Yu.~S.~Kivshar, D.~E.~Pelinovsky, T.~Cretegny, and M.~Peyrard, Phys. Rev. Lett. {\bf 80}, 5032 (1998).

\bibitem{phi6a} V.~A. Gani, A.~E. Kudryavtsev, and M.A. Lizunova, Phys. Rev. D {\bf 89}, 125009 (2014).

\bibitem{phi6b} P.~Dorey, K.~Mersh, T.~Romanczukiewicz, and Y.~Shnir, Phys. Rev. Lett. 107, 091602 (2011).

\bibitem{phi8} V.~A. Gani, V. Lensky, and M.~A. Lizunova, J. High Energy Phys. {\bf 2015} (8), 147 (2015).
    
\bibitem{Khare} A.~Khare, I.~C.~Christov, and A.~Saxena, Phys. Rev. E {\bf 90}, 023208 (2014).

\bibitem{Javidan} K.~Javidan, Phys. Rev. E {\bf 78}, 046607 (2008).
    
\bibitem{DanialPRD} D.~Saadatmand, S.~V.~Dmitriev, and P.~G.~Kevrekidis, Phys. Rev. D {\bf 92}, 056005 (2015).

\bibitem{PREcollisions} S.~V.~Dmitriev, P.~G.~Kevrekidis, and Yu.~S.~Kivshar, Phys. Rev. E {\bf 78}, 046604 (2008).


\end{thebibliography}
\end{document}